\documentclass[twocolumn,showpacs,preprintnumbers,amsmath,amssymb]{revtex4}
\usepackage{amsmath,amssymb,graphicx}
\usepackage{graphicx}
\usepackage{dcolumn}
\usepackage{bm}
\usepackage{epsfig}
\usepackage{here}
\usepackage{float}
\usepackage{amsmath}
\usepackage[usenames]{color}
\usepackage{amsmath}
\usepackage{epsfig}
\usepackage{color}
\usepackage{ulem}

\newcommand{\bea}{\begin{eqnarray}}
\newcommand{\eea}{\end{eqnarray}}

\newcommand{\lred}{}

\begin{document}
\title{Axion as a fuzzy-dark-matter candidate: Proofs in different gauges}
\author{Jai-chan Hwang${}^{1,2}$, Hyerim Noh${}^{3}$}
\address{${}^{1}$Department of Astronomy and Atmospheric
         Sciences, Kyungpook National University, Daegu 702-701, Republic of Korea
         \\
         ${}^{2}$Center for Theoretical Physics of the Universe,
         Institute for Basic Science (IBS), Daejeon, 34051, Republic of Korea
         \\
         ${}^{3}$Theoretical Astrophysics Group, Korea Astronomy and Space Science Institute, Daejeon, Republic of Korea
         }

\date{\today}

\begin{abstract}

Axion as a coherently oscillating massive scalar field is known to behave as a zero-pressure irrotational fluid with characteristic quantum stress on a small scale. In relativistic perturbation theory, the case was proved in the axion-comoving gauge up to fully nonlinear and exact order. Our basic assumption is that the field is oscillating with  Compton frequency and the Compton wavelength is smaller than the horizon scale. Here, we revisit the relativistic proof to the linear order in the other gauge conditions. We show that {\it the same} equation for density perturbation known in the non-relativistic treatment can be derived in two additional gauge conditions: the zero-shear gauge and the uniform-curvature gauge. \lred{The uniform-expansion gauge fails to get the aimed equation, and the quantum stress term is missing in the synchronous gauge.
For comparison, we} present the relativistic density perturbation equations in the zero-pressure {\it fluid} in these gauge conditions. Except for the comoving and the synchronous gauge, the equations strikingly differ from the axion case. We clarify that the relativistic analysis \lred{based on time averaging} is valid for scales {\it larger} than the Compton wavelength\lred{. Below the Compton wavelength, the field is not oscillating, and our oscillatory ansatz does not apply. We suggest an equation valid in all scales in the comoving gauge.} For comparison, we review the non-relativistic quantum hydrodynamics and present the Schr\"odinger equation to first-order post-Newtonian expansion in the cosmological context.

\end{abstract}
\maketitle

%
%
%
\section{Introduction}
                                   \label{sec:Introduction}

A coherently oscillating massive scalar field without interaction is known to behave as a pressureless fluid. An example is an axion whose pseudo nature does not interfere with its cosmological role. Such a scalar field can have a role just after the inflation, with the field oscillating at the bottom of the potential, providing a brief matter-dominated period before the radiation domination. More importantly, it can serve as a cold dark matter \cite{axion-CDM}. Calling a coherently oscillating phase of the massive scalar field as axion, disregarding the mass range of the original QCD axion \cite{Kim-1987}, might be an overuse of the term \cite{Marsh-2016}. Still, here we will continue to use it.

The massive scalar field, in fact, has characteristic stress with quantum origin. The quantum origin is apparent in the non-relativistic treatment based on the fluid formulation of the Schr\"odinger equation known as early as in 1926 \cite{Madelung-1927}, the same year both the Schr\"odinger equation and its relativistic version, the Klein-Gordon equation, appeared; for a historical summary, see Section 1 in \cite{Chavanis-Matos-2017}, and \cite{Dirac-1979}.
We review the non-relativistic treatment based on the Schr\"odinger-Poisson system in Sec.\ \ref{sec:NR}. The quantum stress term with extreme-light mass has recently attracted much attention as the fuzzy dark matter \cite{Hu-Barkana-Gruzinov-2000} enabling to resolve the small-scale tensions encountered in the conventional cold dark matter scenario while enjoying all the success of the cold dark matter in the large scale; for reviews, see \cite{FDM-review}.

The relativistic cosmological perturbation theory is based on Einstein's equation combined with the Klein-Gordon equation in the homogeneous and isotropic cosmological background. The relativistic treatment depends on the gauge choice, especially the temporal one often called the hypersurface or slicing condition; the spatial gauge condition is trivial and unique in the homogeneous and isotropic background \cite{Bardeen-1988}. Previous analyses of the axion perturbation used the zero-shear gauge \cite{Nambu-Sasaki-1990}, the synchronous gauge \cite{Ratra-1991}, the uniform-curvature gauge \cite{Hwang-1997}, and the comoving gauge \cite{Hwang-Noh-2009}. But the density perturbation equation has appeared only in the comoving gauge \cite{Hwang-Noh-2009}.

Here, we present the density perturbation equations in these gauge conditions with an addition of the uniform-expansion gauge. We will show that the proof is possible in the zero-shear gauge and the uniform-curvature gauge beside the already known comoving gauge. The quantum stress term is missing in the synchronous gauge, and the aimed equation does not appear in the uniform-expansion gauge, see Sec.\ \ref{sec:relativistic}\lred{; for the missing quantum stress in the synchronous gauge and an alternate possibility, see one paragraph below Eq.\ (\ref{UEG}) and below Eq.\ (\ref{delta-eq-SG}).} The cases are curious because, except for the comoving gauge and the synchronous gauge, the density perturbation equation of a zero-pressure fluid in the other three gauge conditions looks quite complicated, especially in the super-horizon scale. We compare the axion density perturbation equations with the ones in the zero-pressure fluid in Sec.\ \ref{sec:zero-pressure}.

We consider a massive scalar field, and our proof is based on the oscillating field {\it ansatz} assuming the Compton wavelength smaller than the horizon scale. In the relativistic analysis, we find the consistency of full equations demands the Compton wavelength smaller than the scale we are interested in; \lred{this is because, in the sub-Compton scale, the scalar field is not oscillatory. By considering the fluid behavior of the scalar field in such a regime in the comoving gauge, we suggest a density perturbation equation valid in {\it all} scales in the same gauge condition, see Eq.\ (\ref{delta-eq-axion}).}

Our relativistic perturbation theory in Sec.\ \ref{sec:relativistic} is confined to the linear order in perturbation and ignores the interaction term. In contrast, the non-relativistic analysis reviewed in Sec.\ \ref{sec:NR} and the post-Newtonian (PN) treatment in the Appendix \ref{sec:Appendix} are presented in an exact form with full nonlinearity and including the interaction terms in the scalar field potential.

%
%
%
\section{Non-relativistic quantum hydrodynamics}
                                \label{sec:NR}

From the Schr\"odinger equation combined with the Poisson's equation, we have
\bea
   & & i \hbar {\partial \psi \over \partial t}
       = - {\hbar^2 \over 2 m} \Delta \psi
       + \sum_{n = 1, 2 \dots} g_{n+1} |\psi|^{2n}\psi
       + m \Phi \psi,
   \label{Schrodinger-eq} \\
   & & \Delta \Phi = 4 \pi G m | \psi |^2
       - \Lambda c^2,
   \label{Poisson-eq}
\eea
where for $n = 1$, $g_2 \equiv 4 \pi \ell_s \hbar^2 / m$ with $\ell_s$ the s-wave scattering length \cite{Dalfovo-etal-1999, Pitaevskii-Stringari-2003}. Equation (\ref{Schrodinger-eq}) with the interaction term up to $n = 1$ is often known as the nonlinear Schr\"odinger equation or the Gross-Pitaevskii equation \cite{Gross-1961, Pitaevskii-1961}. This will be {\it derived} in the Appendix as the non-relativistic ($c \rightarrow \infty$) or zeroth-order post-Newtonian (0PN), i.e., Newtonian, limit of the Klein-Gordon equation; there, we derive the Schr\"odinger equation to 1PN order in the context of cosmological background, see Eq.\ (\ref{Schrodinger-1PN}).
$\Phi$ is the Newtonian gravitational potential with
\bea
   & & g_{00} = - \left( 1 + 2 {\Phi \over c^2} \right),
\eea
and $\Lambda$ is the cosmological constant. The Poisson's equation follows from the Einstein equation in the 0PN ($c \rightarrow \infty$) limit, see the Appendix for derivation. Equation (\ref{Poisson-eq}) can be written as
\bea
   & & \Delta \Phi = 4 \pi G m \left( | \psi |^2
       - | \psi_b |^2 \right),
   \label{Poisson-eq-1}
\eea
with $\psi_b$ the homogeneous background wave function.

Using the Madelung transformation \cite{Madelung-1927}
\bea
   & & \psi \equiv \sqrt{\varrho \over m} e^{i m u/\hbar},
   \label{M-transformation}
\eea
with $\varrho$ and $u$ interpreted as the density and velocity potential, respectively, the imaginary and real parts of the Schr\"odinger equation give
\bea
   & & \dot \varrho = - \left( \varrho u^{,i} \right)_{,i},
   \\
   & & \dot u + {1 \over 2} u^{,i} u_{,i}
       = {\hbar^2 \over 2 m^2}
       {\Delta \sqrt{\varrho} \over \sqrt{\varrho}}
       - \sum_{n = 1, 2 \dots} {g_{n+1} \over m}
       \left( {\varrho \over m} \right)^n
   \nonumber \\
   & & \qquad
       - \Phi.
   \label{Euler-u}
\eea
These can be arranged as the continuity and Euler equations, respectively. Together with the Poisson equation, we have
\bea
   & & \dot \varrho
       + \nabla \cdot \left( \varrho {\bf u} \right)
       = 0,
   \label{continuity-Q} \\
   & & \dot {\bf u} + {\bf u} \cdot \nabla {\bf u}
       = {\hbar^2 \over 2 m^2} \nabla
       \left( {\Delta \sqrt{\varrho} \over \sqrt{\varrho}} \right)
       - \sum_{n = 1, 2 \dots} {g_{n+1} \over m^{n+1}}
       \nabla \varrho^n
   \nonumber \\
   & & \qquad
       - \nabla \Phi,
   \label{Euler-Q} \\
   & & \Delta \Phi = 4 \pi G \varrho
   - \Lambda c^2,
   \label{Poisson-Q}
\eea
where we introduce ${\bf u} \equiv \nabla u$, thus the flow vector ${\bf u}$ is irrotational. The first term in the right-hand side of Eq.\ (\ref{Euler-Q}) is the characteristic quantum stress appearing in the Euler equation. The fuzzy nature as the dark matter is played by this term which we may call the Madelung term. The second term is an interaction pressure caused by the nonlinear interaction term. Compared with the Newtonian fluid equation \cite{Hwang-Noh-Puetzfeld-2008}
\bea
   & & \dot u_i + u^j \nabla_j u_i
       = - {1 \over \varrho} \left( \nabla_j \Pi_i^j
       + \nabla_i p \right)
       - \nabla_i \Phi,
   \label{Euler-fluid}
\eea
with the pressure (isotropic stress) $p$ and the anisotropic stress $\Pi_{ij}$ (with $\Pi^i_i \equiv 0$), we have \cite{Takabayasi-1952}
\bea
   & & \Pi^{\rm Q}_{ij} = - {\hbar^2 \over 4 m^2}
       \varrho \left( \nabla_i \nabla_j
       - {1 \over 3} \delta_{ij} \Delta \right) \ln{\varrho},
   \nonumber \\
   & & p^{\rm Q}
       = - {\hbar^2 \over 12 m^2} \varrho \Delta \ln{\varrho}
       + \sum_{n = 1, 2 \dots} {n g_{n+1} \over n + 1}
       \left( {\varrho \over m} \right)^{n+1},
   \label{stress}
\eea
thus the Madelung term has both isotropic and anisotropic stresses, justifying calling it the {\it quantum stress} as correctly pointed out in \cite{Takabayasi-1952}.

Equations (\ref{continuity-Q})-(\ref{Poisson-Q}) guarantee that $\varrho$ and $u$ defined in Eq.\ (\ref{M-transformation}) can be identified as the density and velocity potential, respectively. It is important to notice, however, that these identifications apply {\it only} to the non-relativistic limit \cite{Chavanis-Matos-2017}, see below Eq.\ (\ref{T_ab-0PN}). The proper identification of the fluid quantities should be made based on decomposition of the energy-momentum tensor using the four-vector; see Eq.\ (\ref{Tab-identification}), and Eq.\ (\ref{Tab-identification-SF}) to the linear order perturbation in cosmology.

The equivalence between the two systems, the Schr\"odinger equation in (\ref{Schrodinger-eq}) compared with its quantum-fluid formulation in Eqs.\ (\ref{continuity-Q}) and (\ref{Euler-Q}), in the absence of the interaction term, was challenged in \cite{Wallstrom-1994}. These two are {\it not} equivalent. An important difference appears in the presence of quantized circulation (or vortex, $\nabla \times {\bf v}$) in the Schr\"odinger system, as is well known in superfluids and Bose-Einstein condensates \cite{Onsager-1949, Takabayasi-1952, Feynman-1955, Gross-1961, Pitaevskii-1961, Lifshitz-Pitaevskii-1980, Wallstrom-1994, Pethick-Smith-2002, Pitaevskii-Stringari-2003, Tsubota-etal-2013, Barenghi-Parker-2016}. The difference is reflected in the simulations based on the Schr\"odinger formulation, with the interference pattern and quantized vortex \cite{Hopkins-2019, FDM-review}. These wave-like features are not available in the simulations based on the fluid formulation \cite{Zhang-Liu-Chi-2019}; see, however, \cite{Li-Hui-Bryan-2019}.

\subsection{Cosmological perturbation}

In the cosmological context, we introduce the spatially homogeneous and isotropic background and perturbation
\bea
   & & \varrho \rightarrow \varrho + \delta \varrho, \quad
       {\bf u} = H {\bf r} + {\bf v},
   \label{pert-N}
\eea
where $H \equiv {\dot a / a}$ with $a(t)$ the cosmic scale factor; we set $\delta \varrho \equiv \varrho \delta$, and $\Phi$ has a perturbed part {\it only}. To the background order, Eqs.\ (\ref{continuity-Q})-(\ref{Poisson-Q}) give
\bea
   & & \dot \varrho + 3 H \varrho = 0, \quad
       {\ddot a \over a} = - {4 \pi G \over 3} \varrho
       + {\Lambda c^2 \over 3}.
   \label{BG-N}
\eea
These are the Friedmann equations in the absence of pressure and background curvature. Moving to the comoving coordinate ${\bf x}$ where
\bea
   & & {\bf r} \equiv a (t) {\bf x},
\eea
thus
\bea
   & & \nabla_{\bf r}
       = {1 \over a} \nabla_{\bf x},
   \\
   & & {\partial \over \partial t} \bigg|_{\bf r}
       = {\partial \over \partial t} \bigg|_{\bf x}
       + \left( {\partial \over \partial t} \bigg|_{\bf r}
       {\bf x} \right) \cdot \nabla_{\bf x}
       = {\partial \over \partial t} \bigg|_{\bf x}
       - H {\bf x} \cdot \nabla_{\bf x},
   \nonumber
\eea
neglecting the subindex ${\bf x}$, Eqs.\ (\ref{continuity-Q})-(\ref{Poisson-Q}) become \cite{Chavanis-2012}
\bea
   & & \dot \delta + {1 \over a} \nabla \cdot
       \left[ \left( 1 + \delta \right) {\bf v} \right] = 0,
   \label{continuity-pert-N} \\
   & & \dot {\bf v} + H {\bf v}
       + {1 \over a} {\bf v} \cdot \nabla {\bf v}
       = {\hbar^2 \over 2 m^2} {1 \over a^3} \nabla
       \left( {\Delta \sqrt{1 + \delta} \over \sqrt{1 + \delta}} \right)
   \nonumber \\
   & & \qquad
       - \sum_{n = 1, 2 \dots} {g_{n+1} \over m^{n+1}}
       {1 \over a} \nabla [\varrho (1 + \delta)]^n
       - {1 \over a} \nabla \Phi,
   \label{Euler-pert-N} \\
   & & {\Delta \over a^2} \Phi = 4 \pi G \varrho \delta.
   \label{Poisson-pert-N}
\eea
Combining these, we have
\bea
   & & \ddot \delta + 2 H \dot \delta
       - 4 \pi G \varrho \delta
       + {\hbar^2 \over 2 m^2}
       {\Delta \over a^4} \left(
       {\Delta \sqrt{1 + \delta} \over \sqrt{1 + \delta}} \right)
   \nonumber \\
   & & \qquad
       - \sum_{n = 1, 2 \dots} {g_{n+1} \over m^{n+1}}
       {\Delta \over a^2} [\varrho (1 + \delta)]^n
   \nonumber \\
   & & \qquad
       = {1 \over a^2} \nabla \cdot
       \left( {\bf v} \cdot \nabla {\bf v} \right)
       - {1 \over a^2} \left[
       a \nabla \cdot \left( \delta {\bf v} \right)
       \right]^{\displaystyle{\cdot}}.
\eea
These are valid to fully nonlinear order in perturbation. Ignoring the quantum stress and the interaction pressure terms, this equation can be derived to the second-order perturbation in relativistic perturbation theory in the comoving gauge, see Eq.\ (342) in \cite{Noh-Hwang-2004}. Newtonian hydrodynamic equations in the absence of pressure are closed to the second-order, which is not the case in Einstein's gravity. Thus, all higher-order perturbations in Einstein's gravity are pure relativistic corrections, see Section 5 in \cite{Hwang-Noh-2006} to the third-order perturbation, and Section 5.1 in \cite{Hwang-Noh-2013} to the fully nonlinear and exact order perturbation in Einstein's gravity. These relativistic corrections were derived in the comoving gauge. As we will show in Sec.\ \ref{sec:zero-pressure}, in other gauges the equations are more involved even to the linear order perturbation, especially in the super-horizon scale.

To the linear order, we have
\bea
   & & \ddot \delta + 2 H \dot \delta
       - 4 \pi G \varrho \delta
       + {\hbar^2 \Delta^2 \over 4 m^2 a^4} \delta
   \nonumber \\
   & & \qquad
       - \sum_{n = 1, 2 \dots} {g_{n+1} \over m^{n+1}}
       n \varrho^n {\Delta \over a^2} \delta
       = 0.
   \label{delta-eq-NR}
\eea
In the absence of the interaction pressure term, this is the density perturbation equation for axion in the non-relativistic limit; our relativistic analysis in the next section will show that {\it the same} equation is valid in the relativistic analysis in three gauge conditions, see Eq.\ (\ref{delta-eq-axion}).

In perturbation theory, the interaction term is troublesome: to the background order, the presence of interaction term in Eq.\ (\ref{Euler-u}) is difficult to interpret. Similarly, in relativistic perturbation theory in Sec.\ \ref{sec:relativistic}, the interaction terms may cause breakdown of the coherent oscillation in the background, see below Eq.\ (\ref{V}).

Compared with the relativistic analysis where rapid oscillation of the massive scalar field is used as an {\it ansatz} in Eq.\ (\ref{ansatz}) in the Klein-Gordon equation, instead of the Schr\"odinger equation, in non-relativistic analysis, the oscillatory nature is used in the relation between the field ($\phi$) and the wave function ($\psi$) in Eqs.\ (\ref{Klein-transformation}) or (\ref{K-transformation}); Eqs.\ (\ref{K-transformation}) and (\ref{ansatz}) are in fact {\it the same} only using different notations.

\subsection{Cosmological context}

In the cosmological context, in the zeroth-order post-Newtonian (0PN) approximation, or equivalently in the  non-relativistic limit ($c \rightarrow \infty$), we have the Schr\"odinger equation and the Poisson equation in expanding medium
\bea
   & & i \hbar \left( \dot \psi
       + {3 \over 2} H \psi \right)
       = - {\hbar^2 \over 2 m} {\Delta \over a^2} \psi
       + \sum_{n = 1, 2 \dots} g_{n+1} | \psi |^{2n} \psi
   \nonumber \\
   & & \qquad
       + m \Phi \psi,
   \label{Schrodinger-C} \\
   & & {\Delta \over a^2} \Phi
       = 4 \pi G m | \psi |^2 + 3 {\ddot a \over a} - \Lambda c^2.
   \label{Poisson-0PN-C}
\eea
Equation (\ref{Schrodinger-C}) is {\it derived} in the Appendix where we also present the fully relativistic version and the one valid to 1PN order, see Eq.\ (\ref{Schrodinger-GR}) and (\ref{Schrodinger-1PN}), respectively. These follow from the Klein-Gordon equation using the Klein transformation in Eq.\ (\ref{Klein-transformation}). Equation (\ref{Poisson-0PN-C}) is also {\it derived} in the Appendix. In the non-expanding background we recover Eqs.\ (\ref{Schrodinger-eq}) and (\ref{Poisson-eq}) by setting $a \equiv 1$. Equation (\ref{Poisson-0PN-C}) can also be written as
\bea
   & & {\Delta \over a^2} \Phi
       = 4 \pi G m \left( | \psi |^2
       - | \psi_b |^2 \right),
   \label{Poisson-0PN-C-1}
\eea
with $\psi_b$ the homogeneous background wave function.

Under the Madelung transformation in Eq.\ (\ref{M-transformation}), from imaginary and real parts, respectively, we have
\bea
   & & \dot \varrho + 3 H \varrho
       + {1 \over a^2} \left( \varrho u^{,i} \right)_{,i}
       = 0,
   \\
   & & \dot u + {1 \over 2 a^2} u^{,i} u_{,i}
       = {\hbar^2 \over 2 m^2} {1 \over a^2}
       {\Delta \sqrt{\varrho} \over \sqrt{\varrho}}
       - \sum_{n = 1, 2 \dots} {g_{n+1} \over m}
       \left( {\varrho \over m} \right)^n
   \nonumber \\
   & & \qquad
       - \Phi.
\eea
Identifying ${\bf v} \equiv {1 \over a} \nabla u$, we have
\bea
   & & \dot \varrho + 3 H \varrho
       + {1 \over a} \nabla \cdot \left( \varrho {\bf v} \right)
       = 0,
   \label{Continuity-C} \\
   & & \dot {\bf v} + H {\bf v}
       + {1 \over a} {\bf v} \cdot \nabla {\bf v}
       = {\hbar^2 \over 2 m^2} {1 \over a^3} \nabla \left(
       {\Delta \sqrt{\varrho} \over \sqrt{\varrho}} \right)
   \nonumber \\
   & & \qquad
       - \sum_{n = 1, 2 \dots} {g_{n+1} \over m^{n+1}}
       {1 \over a} \nabla [\varrho (1 + \delta)]^n
       - {1 \over a} \nabla \Phi,
   \label{Euler-C} \\
   & & {\Delta \over a^2} \Phi
       = 4 \pi G \varrho + 3 {\ddot a \over a} - \Lambda c^2.
   \label{Poisson-C}
\eea
Using perturbation expansion in Eq.\ (\ref{pert-N}), we recover Eq.\ (\ref{BG-N}) to the background order, and Eqs.\ (\ref{continuity-pert-N})-(\ref{Poisson-pert-N}) to the nonlinear perturbations. Subtracting the background, Eq.\ (\ref{Poisson-C}) becomes
\bea
   & & {\Delta \over a^2} \Phi
       = 4 \pi G \left( \varrho - \varrho_b \right),
\eea
with $\varrho_b$ the homogeneous background density.

The Madelung transformation can be inverted to give
\bea
   & & \varrho = m | \psi |^2, \quad
       {\bf v} = {\hbar \over 2 i m a} \left(
       {\nabla \psi \over \psi}
       - {\nabla \psi^* \over \psi^*} \right).
\eea
Using this we can directly show that Eqs.\ (\ref{Continuity-C}) and (\ref{Euler-C}) are valid by Eq.\ (\ref{Schrodinger-C}).

\lred{The non-relativistic analysis in this section began with the Schr\"odinger equation derived from the Klein-Gordon equation using the Klein transformation in Eq.\ (\ref{Klein-transformation}), which separates the rapidly oscillating part with Compton frequency. However, in the sub-Compton scale, the scalar field is not oscillatory, and the whole formulation in this section and Appendix \ref{sec:Appendix} does not apply. For a suggestion to extend the fluid formulation to the sub-Compton scale in a particular gauge condition, see below Eq.\ (\ref{additional-condition}).}

%
%
%
\section{Relativistic analysis}
                                \label{sec:relativistic}

Now, we present the relativistic counterpart of the previous section to the linear order perturbation. Fully relativistic quantum hydrodynamics will be presented on a later occasion \cite{Hwang-Noh-2021}, and here, we consider only the linear perturbation analysis in the cosmological context. Instead of the Schr\"odinger equation with Newtonian gravity represented by the Poisson's equation, in Eqs.\ (\ref{Schrodinger-eq}) and (\ref{Poisson-eq}), or Eqs.\ (\ref{Schrodinger-C}) and (\ref{Poisson-0PN-C}), now we use the Klein-Gordon equation with Einstein's gravity modified by the cosmological constant
\bea
   & & \Box \phi = V_{,\phi},
   \label{KG-eq} \\
   & & R_{ab} - {1 \over 2} g_{ab} R + \Lambda g_{ab}
       = {8 \pi G \over c^4} T_{ab}.
   \label{Einstein-eq}
\eea
Our convention in Lagrangian density is
\bea
   & & L = {c^4 \over 16 \pi G}
       \left( R - 2 \Lambda \right)
       - {1 \over 2} \phi^{;c} \phi_{,c} - V (\phi).
\eea
The energy-momentum tensor is
\bea
   & & T^a_b
       = \phi^{;a} \phi_{,b}
       - \left( {1 \over 2} \phi^{;c} \phi_{,c} + V
       \right) \delta^a_b.
   \label{Tab}
\eea
For a massive scalar field with interaction, we have
\bea
   & & V = {1 \over 2} {m^2 c^2 \over \hbar^2} \phi^2
       + \sum_{n = 1, 2 \dots} {g_{n+1} \over n+1}
       \left( {m \over \hbar^2} \phi^2 \right)^{n + 1}.
   \label{V}
\eea
In this section, we will consider a massive scalar field {\it without} interaction. With the interaction terms, our basic ansatz in Eq.\ (\ref{ansatz}) does {\it not} apply to the background order; i.e., the field no longer coherently oscillates, see \cite{Turner-1983}.

The Schr\"odinger equation in Eqs.\ (\ref{Schrodinger-eq}) and (\ref{Schrodinger-C}) follows from the Klein-Gordon equation in (\ref{KG-eq}), under the Klein transformation \cite{Klein-1927}
\bea
   & & \phi = {\hbar \over \sqrt{m}} \psi e^{-i m c^2 t/\hbar},
   \label{Klein-transformation}
\eea
as the non-relativistic limit ($c \rightarrow \infty$). This can be done regarding $\phi$ as if it is a complex field. As we consider a {\it real} scalar field $\phi$ and a complex wave function $\psi$, however, a more proper way is to expand as
\bea
   & & \phi = {\hbar \over \sqrt{2 m}}
       \left( \psi e^{-i m c^2 t/\hbar}
       + \psi^* e^{+i m c^2 t/\hbar} \right).
   \label{K-transformation}
\eea
Both methods give the same answer, see below Eq.\ (\ref{Schrodinger-GR}) and below Eq.\ (\ref{Tab-MSF-psi}).

\subsection{Cosmological perturbation}
                                         \label{sec:CP}

We consider a {\it flat} Friedmann model supported by a minimally coupled massive scalar field. As the scalar field does not support the (transverse) vector and (transverse-tracefree) tensor type perturbations to the linear order, we consider only the scalar-type perturbation. Our metric convention is \cite{Bardeen-1988, Hwang-Noh-2013}
\bea
   & & g_{00}
       = - a^2 \left( 1 + 2 \alpha \right), \quad
       g_{0i} = - a \chi_{,i},
   \nonumber \\
   & &
       g_{ij} = a^2 \left( 1 + 2 \varphi \right)
       \delta_{ij},
\eea
where $x^0 = \eta$ with $c dt \equiv a d \eta$; we imposed a spatial gauge condition under which all remaining perturbation variables are spatially gauge-invariant \cite{Bardeen-1988}.

The fluid quantities are identified based on the time-like four-vector $u_a$ with $u^a u_a \equiv -1$. In the energy frame, setting the flux four-vector $q_a \equiv 0$, we have \cite{Ellis-1971, Hwang-Noh-2013}
\bea
   & & T_{ab} = \mu u_a u_b
       + p \left( g_{ab} + u_a u_b \right)
       + \pi_{ab},
   \label{Tab-identification}
\eea
with $\pi_{ab} u^b = 0 = \pi^a_a$.

To the linear order, with $\phi \rightarrow \phi + \delta \phi$, \lred{$\mu \rightarrow \mu + \delta \mu$, $p \rightarrow p + \delta p$,} $u_i \equiv a v_i/c$ and $\pi_{ij} \equiv a^2 \Pi_{ij}$, with indices of $v_i$ and $\Pi_{ij}$ raised and lowered using $\delta_{ij}$ as the metric, Eq.\ (\ref{Tab}) gives
\bea
   T^0_0
   &=& - {1 \over 2 c^2} \left( \dot \phi^2
       + {m^2 c^4 \over \hbar^2} \phi^2 \right)
   \nonumber \\
   & &
       - {1 \over c^2} \left( \dot \phi \delta \dot \phi
       - \dot \phi^2 \alpha
       + {m^2 c^4 \over \hbar^2} \phi \delta \phi \right)
   \nonumber \\
   &\equiv& - \mu - \delta \mu,
   \nonumber \\
   T^0_i
   &=& - {1 \over a c} \dot \phi
       \delta \phi_{,i}
       \equiv - {1 \over c} \left( \mu + p \right) v_{,i},
   \nonumber \\
   T^i_j
   &=& \bigg[
       {1 \over 2 c^2} \left( \dot \phi^2
       - {m^2 c^4 \over \hbar^2} \phi^2 \right)
   \nonumber \\
   & &
       + {1 \over c^2} \left( \dot \phi \delta \dot \phi
       - \dot \phi^2 \alpha
       - {m^2 c^4 \over \hbar^2} \phi \delta \phi \right)
       \bigg] \delta^i_j
   \nonumber \\
   &\equiv& \left( p + \delta p \right) \delta^i_j
       + \Pi^i_j,
   \label{Tab-identification-SF}
\eea
where we decompose $v_i \equiv - v_{,i}$ for scalar type perturbation \cite{Noh-Hwang-2004}.
Thus, we read perturbed fluid quantities as
\bea
   & & \delta \mu
       = {1 \over c^2} \left( \dot \phi \delta \dot \phi
       - \dot \phi^2 \alpha
       + {m^2 c^4 \over \hbar^2} \phi \delta \phi \right),
   \nonumber \\
   & &
       \delta p
       = {1 \over c^2} \left( \dot \phi \delta \dot \phi
       - \dot \phi^2 \alpha
       - {m^2 c^4 \over \hbar^2} \phi \delta \phi \right),
   \nonumber \\
   & &
       \left( \mu + p \right) v
       = {1 \over a} \dot \phi
       \delta \phi,
   \label{fluid-MSF-pert}
\eea
and vanishing anisotropic stress, $\Pi^i_j$; this shows that the scalar field does not support the vector and tensor type perturbations.

To the background order, we have the Friedmann equations and the equation of motion
\bea
   & & H^2 = {8 \pi G \over 3 c^2} \mu
       + {\Lambda c^2 \over 3}, \quad
       \dot H = - {4 \pi G \over c^2} \left( \mu + p \right),
   \\
   & & \ddot \phi + 3 H \dot \phi + {m^2 c^4 \over \hbar^2} \phi
       = 0,
   \label{EOM-BG}
\eea
with fluid quantities identified in Eq.\ (\ref{Tab-identification-SF}) as
\bea
   & & \mu = {1 \over 2 c^2} \left( \dot \phi^2
       + {m^2 c^4 \over \hbar^2} \phi^2 \right),
   \nonumber \\
   & &
       p = {1 \over 2 c^2} \left( \dot \phi^2
       - {m^2 c^4 \over \hbar^2} \phi^2 \right).
   \label{fluid-MSF-BG}
\eea

To linear order in perturbation, the basic equations for the scalar-type perturbation, without imposing the temporal gauge condition, are \cite{Bardeen-1988, Noh-Hwang-2004, Hwang-Noh-2013}
\bea
   & & \kappa \equiv 3 H \alpha - 3 \dot \varphi
       - c {\Delta \over a^{2}} \chi,
   \label{eq1} \\
   & & {4 \pi G \over c^2} \delta \mu + H \kappa
       + c^2 {\Delta \over a^{2}} \varphi = 0,
   \label{eq2} \\
   & & \kappa + c {\Delta \over a^{2}} \chi
       - {12 \pi G \over c^4} \left(
       \mu + p \right) a v = 0,
   \label{eq3} \\
   & & \dot \kappa + 2 H \kappa
       + \left( 3 \dot H + c^2 {\Delta \over a^{2}} \right)
       \alpha = {4 \pi G \over c^2} \left( \delta \mu
       + 3 \delta p \right),
   \label{eq4} \\
   & & \varphi + \alpha
       - {1 \over c} \left( \dot \chi + H \chi \right)
       = 0,
   \label{eq5} \\
   & & \delta \dot \mu + 3 H \left( \delta \mu
       + \delta p \right)
       = \left( \mu + p \right)
       \left( \kappa - 3 H \alpha
       + {\Delta \over a} v \right),
   \label{eq6} \\
   & & {1 \over a^{4}} \left[ a^4 \left( \mu
       + p \right) v
       \right]^{\displaystyle\cdot}
       = {c^2 \over a}\left[ \delta p
       + \left( \mu + p \right) \alpha \right].
    \label{eq7}
\eea
These are Einstein's equation and the energy-momentum conservation equation, thus a redundant set. The equation of motion gives
\bea
   & & \delta \ddot \phi + 3 H \delta \dot \phi
       - c^2 {\Delta \over a^2} \delta \phi
       + {m^2 c^4 \over \hbar^2} \delta \phi
   \nonumber \\
   & & \qquad
       = \dot \phi \left( \kappa + \dot \alpha \right)
       + \left( 2 \ddot \phi + 3 H \dot \phi \right) \alpha.
   \label{EOM-pert}
\eea
Using Eqs.\ (\ref{fluid-MSF-pert}) and (\ref{fluid-MSF-BG}), this also follows from Eq.\ (\ref{eq6}), and Eq.\ (\ref{eq7}) is identically satisfied.

The above set of perturbation equations is presented without imposing the temporal gauge (hypersurface or slicing) condition, and all variables used are spatially gauge invariant. We have the following temporal gauge conditions: the comoving gauge (CG, $v \equiv 0$), the zero-shear gauge (ZSG, $\chi \equiv 0$), the uniform-curvature gauge (UCG, $\varphi \equiv 0$), the uniform-expansion gauge (UEG, $\kappa \equiv 0$), the uniform-density gauge (UDG, $\delta \mu \equiv 0$), the uniform-field gauge (UFG, $\delta \phi \equiv 0$), and the synchronous gauge (SG, $\alpha \equiv 0$). These include most of the gauge conditions used in the literature \cite{Bardeen-1980, Bardeen-1988}. Except for the SG, all the other gauge conditions completely fix the gauge degree of freedom, and each variable in these gauge conditions has a unique gauge-invariant combination of variables. These statements concerning the gauge issue are valid to fully nonlinear order in perturbations \cite{Noh-Hwang-2004, Hwang-Noh-2013}. In a single-component fluid supported by the scalar field, in the ordinary scaler field case the UFG coincides with the CG. As we consider a single component field, our CG is the axion-comoving gauge. However, as $\delta \phi$ rapidly oscillates, and as we will consider time-average over the oscillation, the UFG could differ from the CG in the case of the axion.

\subsection{Axion perturbation}
                                         \label{sec:axion-P}

We take an {\it ansatz} \cite{Ratra-1991}
\bea
   & &\phi (t) + \delta \phi ({\bf x}, t)
       = a^{-3/2} \phi_{+0} \left[ 1
       + \Phi_+ ({\bf x}, t) \right] \sin{(m c^2 t/\hbar)}
   \nonumber \\
   & & \qquad
       + a^{-3/2} \phi_{-0} \left[ 1
       + \Phi_- ({\bf x}, t) \right] \cos{(m c^2 t/\hbar)},
   \label{ansatz}
\eea
which is {\it the same} as the Klein transformation in Eq.\ (\ref{Klein-transformation}).
We will strictly consider {\it only} the leading order in
\bea
   & & {\hbar H \over m c^2}
       = {\lambdabar_c \over \lambda_H}
       = 2.13 \times 10^{-11} {h_{100} \over m_{22}} {H \over H_0},
\eea
with $\lambdabar_c \equiv \lambda_c/(2 \pi)$ and $\lambda_c \equiv h/(mc)$ the Compton wavelength and $\lambda_H \equiv c/H$ the Hubble horizon scale, respectively; we set $H \equiv 100 h_{100} {\rm km}/{\rm sec Mpc}$, $m_{22} \equiv m c^2/(10^{-22} {\rm eV})$ and the index $0$ indicates the present epoch.

Fluid variables in Eqs.\ (\ref{fluid-MSF-pert}) and (\ref{fluid-MSF-BG}) are composed with quadratic combinations of the field variables. We take the time-average over the oscillation, thus for example,
\bea
   & & \left( \mu + p \right) v
       = {1 \over a} \langle \dot \phi \delta \phi \rangle,
\eea
etc., with
\bea
   & & \langle f(t) \rangle \equiv {m \over 2 \pi}
       \int^{2\pi/m}_0 f(t^\prime) d t^\prime.
\eea

To leading order in $\hbar H/(m c^2)$, the background fluid quantities in Equation (\ref{fluid-MSF-BG}) give
\bea
   & & \mu = {m^2 c^2 \over 2 \hbar^2 a^3} \left( \phi_{+0}^2
       + \phi_{-0}^2 \right), \quad
       p = 0.
   \label{fluid-axion-BG}
\eea
Thus, to the background order, the rapidly oscillating massive scalar field behaves as a pressureless fluid. Under our {\it ansatz}, Eq.\ (\ref{EOM-BG}) is also valid to the leading order in $m c^2/(\hbar H)$.

To the perturbed order, Equation (\ref{fluid-MSF-pert}) gives
\bea
   & & {\delta \mu \over \mu}
       = {2 \over \phi_{+0}^2 + \phi_{-0}^2}
       \left( \phi_{+0}^2 \Phi_+
       + \phi_{-0}^2 \Phi_- \right) - \alpha,
   \nonumber \\
   & &
       {\delta p \over \mu} = - \alpha, \quad
       v = {\hbar \over a m} {\phi_{+0} \phi_{-0}
       \over \phi_{+0}^2 + \phi_{-0}^2}
       \left( \Phi_- - \Phi_+ \right).
   \label{fluid-axion}
\eea
Using the axion fluid quantities in Eqs.\ (\ref{fluid-axion-BG}) and (\ref{fluid-axion}), Eqs.\ (\ref{eq1})-(\ref{eq7}) become
\bea
   & & \kappa = 3 H \alpha - 3 \dot \varphi
       - c {\Delta \over a^{2}} \chi,
   \label{eq1-axion} \\
   & & {4 \pi G \over c^2} \delta \mu + H \kappa
       + c^2 {\Delta \over a^{2}} \varphi = 0,
   \label{eq2-axion} \\
   & & \kappa + c {\Delta \over a^{2}} \chi
       = {12 \pi G \over c^4} \mu a v,
   \label{eq3-axion} \\
   & & \dot \kappa + 2 H \kappa
       = {4 \pi G \over c^2} \delta \mu
       - c^2 {\Delta \over a^{2}} \alpha,
   \label{eq4-axion} \\
   & & \varphi + \alpha
       = {1 \over c} \left( \dot \chi + H \chi \right),
   \label{eq5-axion} \\
   & & \dot \delta
       = \kappa + {\Delta \over a} v,
   \label{eq6-axion} \\
   & & {1 \over a} \left( a v \right)^{\displaystyle\cdot} = 0.
   \label{eq7-axion}
\eea
We may set $\mu \equiv {\varrho} c^2$.
As we remarked below Eq.\ (\ref{stress}), ${\varrho}$ in this section {\it differs} from $\varrho$ in the previous section. Here, ${\varrho}$ is the relativistic density defined in the energy-momentum tensor, whereas $\varrho$ in the previous section is defined in Eq.\ (\ref{M-transformation}) and identified with the density in the non-relativistic limit; thus, it is identified as the rest-mass density.

Now, we have the equation of motion. By using the {\it ansatz} in Eq.\ (\ref{ansatz}), the leading $[m c^2/(\hbar H)]^2$ order terms in Eq.\ (\ref{EOM-pert}) cancel, and to the next leading $m c^2/(\hbar H)$ order, the sine and cosine parts, respectively, give
\bea
   & & {2 \over H}
       {\phi_{-0} \over \phi_{+0}} \dot \Phi_-
       + {\hbar \Delta \over m H a^2} \Phi_+
       =
       {\phi_{-0} \over \phi_{+0}} {\kappa \over H}
       + 2 {m c^2 \over \hbar H} \alpha,
   \nonumber \\
   & & {2 \over H}
       {\phi_{+0} \over \phi_{-0}} \dot \Phi_+
       - {\hbar \Delta \over m H a^2} \Phi_-
       =
       {\phi_{+0} \over \phi_{-0}} {\kappa \over H}
       - 2 {m c^2 \over \hbar H} \alpha,
   \label{EOM-axion}
\eea
where we kept ${\hbar \Delta \over m H a^2}$-order terms.

Our task in the following is to derive density perturbation equations in various gauge conditions and check the consistency of the full equations. Equations (\ref{fluid-axion})-(\ref{EOM-axion}) provide the fundamental set of equations for the cosmological axion perturbation without fixing the temporal gauge.

\subsection{Axion density perturbation equation}

For $\kappa \neq 0$, thus {\it excluding} the UEG, from Eqs.\ (\ref{eq4-axion}), (\ref{eq6-axion}) and (\ref{eq7-axion}) we have
\bea
   & & \ddot \delta + 2 H \dot \delta - 4 \pi G \varrho \delta
       = - c^2 {\Delta \over a^2} \alpha.
   \label{delta-eq-alpha}
\eea
The UEG will be treated separately. In the SG, the right-hand-side of Eq.\ (\ref{delta-eq-alpha}) vanishes, thus {\it missing} the quantum stress term. Now, the remaining task is to express $\alpha$ in terms of $\delta$ in the other gauge conditions. This is provided by the equation of motion.

The $v$ and $\delta$ relations in Eq.\ (\ref{fluid-axion}), respectively, using Eq.\ (\ref{eq7-axion}), give
\bea
   & & \dot \Phi_- = \dot \Phi_+, \quad
       2 \dot \Phi_+ = \left( \delta + \alpha
       \right)^{\displaystyle{\cdot}}.
   \label{Phi_pm-relation}
\eea
From Eq.\ (\ref{EOM-axion}), removing the $\kappa$-term and $\alpha$-term, respectively, we have
\bea
   \alpha \lred{= - {\delta p \over \mu}}
       = {1 \over 1 - {\hbar^2 \Delta \over 4 m^2 c^2 a^2}}
       {\hbar^2 \Delta \over 4 m^2 c^2 a^2} \delta, \quad
       2 \dot \Phi_+ = \kappa + {\Delta \over a} v,
   \label{alpha-delta-relation}
\eea
where we used $\delta$ and $v$ relations in Eq.\ (\ref{fluid-axion}), respectively.
We still have not imposed the gauge condition, and Eqs.\ (\ref{Phi_pm-relation}) and (\ref{alpha-delta-relation}) are valid even for the UEG.

Here, we check the consistency with the rest of the equations which we {\it omitted} the our previous work \cite{Hwang-Noh-2009}. In order to have the combined second relations of Eqs.\ (\ref{Phi_pm-relation}) and (\ref{alpha-delta-relation}) consistent with Eq.\ (\ref{eq6-axion}), we need $\alpha \ll \delta$, thus the first relation of Eq.\ (\ref{alpha-delta-relation}) implies
\bea
   & & {\lambda_c^2 \over \lambda^2}
       = {\hbar^2 \over m^2 c^2} {k^2 \over a^2} \ll 1,
   \label{additional-condition}
\eea
where $\lambda = 2 \pi/ k_p \equiv 2 \pi a/k$ is the physical scale of the perturbation; $k$ is the comoving wavenumber with $\Delta \equiv - k^2$. \lred{This happens because the scalar field does not oscillate on the sub-Compton scale, and our ansatz in Eq.\ (\ref{ansatz}) and analysis based on time-average no longer apply. On sub-Compton scale we should go back to the original scalar field equation in Eqs.\ (\ref{fluid-MSF-pert}) and (\ref{EOM-pert}). In Eq.\ (\ref{fluid-MSF-pert}), the axion CG ($v \equiv 0$) is the same as the UFG ($\delta \phi \equiv 0$), and we have $\delta p_v = \delta \mu_v$. Thus, the effective equation of state for the perturbed scalar field in this regime is $\delta p = \delta \mu$ in the CG \cite{Hwang-1991, Hwang-1993, Hwang-1994}. Notice that the perturbed equation of state in Eq.\ (\ref{alpha-delta-relation}) happens to remain valid in the sub-Compton scale. Therefore, as in \cite{Hwang-Noh-2009}, we {\it suggest} that in the CG, the following is valid in {\it all} scales}
\bea
   & & \alpha = \lred{- {\delta p \over \mu} = {1 \over
       1 - {\hbar^2 \Delta \over 4 m^2 c^2 a^2}}
       {\hbar^2 \Delta \over 4 m^2 c^2 a^2} \delta}.
   \label{relations-general}
\eea
\lred{It provides a simple interpolation between the two known asymptotes (the sub-Compton and the super-Compton scales).}

Therefore, except for the SG where $\alpha = 0$, Eq.\ (\ref{delta-eq-alpha}) finally gives
\bea
   \ddot \delta + 2 H \dot \delta - 4 \pi G \varrho \delta
       \lred{
       = {- 1 \over 1 - {\hbar^2 \Delta \over 4 m^2 c^2 a^2}}
       {\hbar^2 \Delta \over 4 m^2 a^2}
       {\Delta \over a^2} \delta.}
   \label{delta-eq-axion}
\eea
\lred{In super-Compton scale, this} is valid for the CG, the ZSG, and the UCG, and {\it coincides} with the non-relativistic one in Eq.\ (\ref{delta-eq-NR})\lred{; in the CG, we suggest that it is valid in {\it all} scales}. By setting the gravity term (the third term in the left-hand-side) equal to the quantum stress we have the quantum Jeans scale, $\lambda_J \equiv {2 \pi a / k_J}$, as
\bea
   \lambda_J
       = \pi \sqrt{\hbar \over m \sqrt{\pi G \varrho}}
       = {55.6 {\rm kpc} \over \sqrt{ m_{22}
       \sqrt{\Omega_{m0}} h_{100} }}
       \left( {\varrho_0 \over \varrho} \right)^{1/4},
\eea
where $\Omega_{m} \equiv 8 \pi G \varrho/(3 H^2)$ is the density parameter of the matter (axion) component.

As we have
\bea
   & & {\lambda_c \over \lambda}
       = {\hbar k_p \over m c}
       = 4.0 \times 10^{-7} {1 \over m_{22}}
       {1 {\rm Mpc} \over \lambda_0} {a_0 \over a},
\eea
$\lambda_c \sim 0.4 {\rm pc}$ \lred{for $m \sim 10^{-22} {\rm eV}$}, and $\lambda_c \sim 12 {\rm cm}$ for a QCD axion with $m \sim 10^{-5} {\rm eV}$.

\subsection{The case of UEG}
                                      \label{sec:UEG}

The UEG sets $\kappa \equiv 0$. Equations (\ref{eq6-axion}) and (\ref{eq7-axion}), and Eq.\ (\ref{eq4-axion}), respectively, give
\bea
   & & \ddot \delta + 2 H \dot \delta = 0, \quad
       4 \pi G \varrho \delta = c^2 {\Delta \over a^2} \alpha.
   \label{UEG-eqs}
\eea
Thus, Eq.\ (\ref{delta-eq-alpha}) is divided into two part, and the UEG fails to deliver an appreciable form of density perturbation equation. We can still proceed the analysis, and show that Eq.\ (\ref{eq3-axion}), (\ref{eq5-axion}) and (\ref{eq7-axion}) give
\bea
   & & \dot \chi + H \chi = 0, \quad
       \varphi = - \alpha.
   \label{UEG}
\eea
Equations (\ref{Phi_pm-relation})-(\ref{relations-general}) remain valid, and we can further check that the full set of equations is {\it consistent}. Thus, although we failed to get a proper density perturbation equation, we can still accept the above (a sort of) segmented equations as the case of axion perturbation in the UEG.

The SG, similarly failed to recover the quantum stress term. In Eq.\ (\ref{relations-general}), the SG does not necessarily imply $\delta = 0$. \lred{As we have $(\lambda_c/\lambda)^2 \delta = - 4 \alpha = 0$, we may have $(\lambda_c/\lambda)^2$ vanishing in the super-Compton scale instead of $\delta$. Thus $\delta$ is undetermined by this relation in the super-Compton scale; still, this relation could be a hint for doubting the absence of quantum stress in the SG, see below eq.~(\ref{delta-eq-SG}).} The SG fails to fix the temporal gauge degree of freedom completely, and the non-vanishing $v$ in Eq.\ (\ref{eq7-axion}) is the gauge mode \cite{Lifshitz-1946}. In order to fix this remnant gauge mode, we can further impose the condition $v = 0$, which is the same as the comoving gauge condition. With this, we can show the consistency of the rest of the equations; in fact, even without the condition on $v$, we can show the consistency. As we set $\alpha \equiv 0$ (which is the source of the quantum stress in other gauge conditions) as the gauge condition, \lred{we have $\delta p = 0$ in Eq.\ (\ref{fluid-axion}), and} the quantum stress term is missing in the SG\lred{; however, as mentioned above, there is room for doubting the absence of the quantum stress and an alternate possibility, see below Eq.\ (\ref{delta-eq-SG}).}

Similarly, we can check the consistency of the complete set of equations in (\ref{fluid-axion})-(\ref{EOM-axion}) for the CG, the ZSG, and the UCG.

Failing to get proper equation known in non-relativistic limit in the UEG and the SG, is not strange or new. For example, the UDG sets $\delta \equiv 0$ as the gauge condition and fails to get a decent density perturbation equation in that gauge. Still, the UDG is a \lred{fine} gauge condition often used in cosmology as $\zeta \equiv \varphi + \delta \mu /[3 (\mu + p)] \equiv \varphi_\delta = \varphi$ in the UDG.

\subsection{Zero-pressure fluid perturbation}
                                     \label{sec:zero-pressure}

Now, we compare our density perturbation equations in various gauge conditions with the ones in the zero-pressure fluid. Besides the absence of the quantum stress term, the density perturbation equations in general {\it differ} from the simple one in Eq.\ (\ref{delta-eq-axion}). {\it Only} for the CG and the SG we have Eq.\ (\ref{delta-eq-axion}) without the quantum stress \cite{Lifshitz-1946, Bardeen-1980}. In the case of the ZSG, the UCG and the UEG, only in the sub-horizon scale with ${c^2 \Delta \over a^2 H^2} \gg 1$ we have Eq.\ (\ref{delta-eq-axion}) \cite{Hwang-Noh-1999}.

In the CG, from Eqs.\ (\ref{eq4}), (\ref{eq6}) and (\ref{eq7}), we can derive
\bea
   & & \ddot \delta_v + 2 H \dot \delta_v
       - 4 \pi G \varrho \delta_v
       = {\Delta \over a^2} {\delta p_v \over \varrho},
   \label{delta-eq-CG}
\eea
where $\delta_v \equiv \delta + 3 H a v/c^2$ is a unique gauge-invariant combination which is {\it the same} as $\delta$ in the comoving gauge setting $v \equiv 0$. Similarly, we have [see Section VI.B. in \cite{Noh-Hwang-2004}]
\bea
   & & \delta_{\alpha} \equiv \delta
       + 3 H \int^t \alpha dt, \quad
       \delta_{\chi} \equiv \delta + 3 H \chi/c,
   \nonumber \\
   & &
       \delta_{\varphi} \equiv \delta + 3 \varphi,\quad
       \delta_{\kappa} \equiv \delta
       - {3 H \over 3 \dot H + c^2 {\Delta \over a^2}} \kappa.
\eea
One exception is the SG where for $\delta_{\alpha}$, the lower-bound of integration gives the remnant gauge mode.

In the SG, from Eqs.\ (\ref{eq4}), (\ref{eq6}) and (\ref{eq7}), we can derive
\bea
   & & \ddot \delta_{\alpha} + 2 H \dot \delta_{\alpha}
       - 4 \pi G \varrho \delta_{\alpha}
       = {\Delta \over a^2} {\delta p_\alpha \over \varrho}.
   \label{delta-eq-SG}
\eea
\lred{For vanishing pressure, the equation is closed to the second order because} the remnant gauge mode for $\delta_{\alpha}$ {\it happens} to behave the same as the decaying solution \cite{Lifshitz-1946}. In Eqs.\ (\ref{delta-eq-CG}) and (\ref{delta-eq-SG}), we kept $\delta p$ to include the \lred{non-relativistic order} pressure perturbation with $\delta p \equiv c_s^2 \delta \varrho \lred{+ e}$ and $c_s^2 \equiv \dot p/\dot \varrho$\lred{; $e \equiv \delta p - c_s^2 \delta \varrho$ is the gauge-invariant entropic perturbation. Notice that Eq.\ (\ref{delta-eq-CG}) in the CG remains valid in the case of the axion. We can recover the pressure perturbation in the SG from the known one in the CG through gauge transformation; let $\delta p_v = v_s^2 \delta \varrho_v$ with known $v_s^2$ as in Eq.\ (\ref{relations-general}) for the axion fluid. Using the gauge-invariance of $e = \delta p_v - c_s^2 \delta \varrho_v$, we have $\delta p_\alpha = \delta p_v + c_s^2 (\delta \varrho_\alpha - \delta \varrho_v)$ and $\delta \varrho_v \equiv \delta \varrho - \dot \varrho a v = \delta \varrho_\alpha - \dot \varrho a v_\alpha$; by the presence of $v_\alpha$-term, Eq.\ (\ref{delta-eq-SG}) is not closed to the second order. In the literature \cite{Hlozek-2015, Poulin-2018}, this method is used to recover the quantum stress in the SG; although we showed $\delta p = 0$ in the SG, potential ambiguity is mentioned in one paragraph below Eq.\ (\ref{UEG}); we can tell which fluid approximation is correct by comparing these to a direct numerical solution of the scalar field perturbation.} In the other gauges, as the terms become complicated, we ignore $\delta p$.

In the ZSG, from Eqs.\ (\ref{eq1})-(\ref{eq7}), we can derive
\bea
   & & \ddot \delta_{\chi} + 2 H \dot \delta_{\chi}
       - 8 \pi G \varrho \delta_{\chi}
       + {3 H^2 + 6 \dot H + {c^2 \Delta \over a^2} \over
       3 H^2 - \left( 1 + {c^2 \Delta \over 12 \pi G \varrho a^2} \right) {c^2 \Delta \over a^2}}
   \nonumber \\
   & & \qquad \times
       \left[ H \dot \delta_{\chi}
       + \left( 1 + {c^2 \Delta \over 12 \pi G \varrho a^2} \right)
       4 \pi G \varrho \delta_{\chi} \right] = 0.
   \label{delta-eq-ZSG}
\eea
In the UCG, from Eqs.\ (\ref{eq1})-(\ref{eq4}), (\ref{eq6}) and (\ref{eq7}), we can derive
\bea
   \ddot \delta_{\varphi}
       + {3 - 2 {c^2 \Delta \over 12 \pi G \varrho a^2} \over
       1 - {c^2 \Delta \over 12 \pi G \varrho a^2}} H \dot \delta_{\varphi}
       + { {c^2 \Delta \over 12 \pi G \varrho a^2} \over
       1 - {c^2 \Delta \over 12 \pi G \varrho a^2}}
       4 \pi G \varrho \delta_{\varphi}
       = 0.
   \label{delta-eq-UCG}
\eea
In the UEG, from Eqs.\ (\ref{eq4}), (\ref{eq6}) and (\ref{eq7}), we can derive
\bea
   & & \ddot \delta_{\kappa} + 2 H \dot \delta_{\kappa}
       - {1 \over a^2} \left(
       {a^2 H \over
       1 - {c^2 \Delta \over 12 \pi G \varrho a^2}}
       \delta_{\kappa} \right)^{\displaystyle{\cdot}}
   \nonumber \\
   & & \qquad
       + { {c^2 \Delta \over 12 \pi G \varrho a^2} \over
       1 - {c^2 \Delta \over 12 \pi G \varrho a^2}}
       4 \pi G \varrho \delta_{\kappa} = 0.
   \label{delta-eq-UEG}
\eea
Similar comparisons between gauges were made in a more general context of including the background curvature in \cite{Hwang-Noh-1999}. Our equations include the cosmological constant.

Notice the striking difference of Eqs.\ (\ref{delta-eq-ZSG}) and (\ref{delta-eq-UCG}) compared with Eq.\ (\ref{delta-eq-CG})\lred{,} and Eq.\ (\ref{delta-eq-UEG}) compared with Eq.\ (\ref{UEG-eqs}). {\it Only} in the sub-horizon limit with ${c^2 \Delta \over a^2 H^2} \gg 1$, Eqs.\ (\ref{delta-eq-ZSG})-(\ref{delta-eq-UEG}) coincide with Eq.\ (\ref{delta-eq-CG}), \cite{Bardeen-1980}.

%
%
%
\section{Discussion}
                                   \label{sec:Discussion}

We presented relativistic derivations of cosmological linear density perturbation equations for a coherently oscillating massive scalar field in several gauge conditions, see Sec.\ \ref{sec:relativistic}. We used the Klein-Gordon equation combined with the Einstein equation in a flat cosmological background. The results depend on the gauge choice, and we show that {\it only} in the CG, the ZSG, and the UCG, we can consistently derive the same equation known in the non-relativistic treatment reviewed in Sec.\ \ref{sec:NR}. The SG {\it fails} to recover the quantum stress term, and the UEG leads to a density perturbation equation segmented into two parts as in Eq.\ (\ref{UEG-eqs}). In the absence of the quantum stress term, the equations coincide with the zero-pressure fluid ones {\it only} in the CG and the SG; the equations in the ZSG, the UCG, and the UEG are more involved, see Eqs.\ (\ref{delta-eq-ZSG})-(\ref{delta-eq-UEG}).

We clarified that \lred{our ansatz} is reliable only for scales larger than the Compton wavelength, see Eq.\ (\ref{additional-condition}). \lred{On scales smaller than the Compton wavelength, the scalar field no longer oscillates, and our ansatz does not apply. In this case, we can go back to the original scalar field equation and show that in the sub-Compton scale, Eq.\ (\ref{relations-general}) happens to remain valid. Thus, we suggested that, in the CG, Eq.\ (\ref{delta-eq-axion}) is valid in {\it all} scales.}

In the relativistic perturbation theory, the non-relativistic nature of the axion was proved to the fully nonlinear and exact order in \cite{Hwang-Noh-Park-2015, Noh-Hwang-Park-2017}. The proof was made in the CG. In the zero-pressure fluid, the analyses in the other gauges are quite complicated as Eqs.\ (\ref{delta-eq-ZSG})-(\ref{delta-eq-UEG}) represent even to the linear order. However, in the axion case, as the linear density perturbation equations coincide exactly in the CG, the ZSG, and the UCG as in Eq.\ (\ref{delta-eq-axion}), similar proof in the ZSG and the UCG might be feasible to the fully nonlinear and exact order. The differences between the axion and the zero-pressure fluid in the gauges other than the CG are remarkable. {\it Only} in the CG, the density perturbation equation of the axion coincides with the one of fluid.

For comparison, we reviewed the non-relativistic derivation in the context of Schr\"odinger equation combined with Poisson equation, see Sec.\ \ref{sec:NR}. Here, the derivation is based on the Madelung transformation of Schr\"odinger equation which leads to the hydrodynamic equations with a characteristic quantum stress term appearing in Euler equation, see Eqs.\ (\ref{Euler-Q}) and (\ref{Euler-C}). In our way to derive the non-relativistic Schr\"odinger equation combined with Newtonian gravity, we presented the Schr\"odinger equation valid to 1PN order in the Appendix. Extension to full 1PN equations of the quantum fluid formulation will be investigated on a later occasion \cite{Hwang-Noh-2021}.

\section*{Acknowledgments}

We wish to thank an anonymous referee for constructive questions and comments. We thank Jiajun Zhang for useful discussion on simulations. H.N.\ was supported by the National Research Foundation (NRF) of Korea funded by the Korean Government (No. 2018R1A2B6002466 and No. 2021R1F1A1045515). J.H.\ was supported by IBS under the project code IBSR018-D1 and by NRF of Korea funded by the Korean Government (No.\ NRF-2019R1A2C1003031).

\begin{widetext}
%
%
%
\appendix

\section{Schr\"odinger equation to 1PN order}
                                   \label{sec:Appendix}

Here we derive the relativistic Schr\"odinger equation to 1PN order and the Poisson equation to 0PN order. Under the Klein transformation in Eq.\ (\ref{Klein-transformation}), with \lred{$\phi^{2n + 1} \rightarrow |\phi|^{2n} \phi$}, the Klein-Gordon equation in Eqs.\ (\ref{KG-eq}) and (\ref{V}) gives
\bea
   & & \Box \phi - {m^2 c^2 \over \hbar^2} \phi
       - \lred{ \sum_{m = 1,2 \dots} 2 g_{n+1}
       {\sqrt{m} \over \hbar}
       \left( {\sqrt{m} \over \hbar} \phi \right)^{2n + 1}}
   \nonumber \\
   & & \qquad
       = {\hbar \over \sqrt{m}} e^{-i m c^2 t/\hbar} \left[ \Box \psi
       - {2 i m \over \hbar} c g^{0c} \psi_{,c}
       + {i m \over \hbar} c g^{ab} \Gamma^0_{ab} \psi
       - {m^2 \over \hbar^2}
       c^2 \left( g^{00} + 1 \right) \psi
       - \sum_{n = 1, 2 \dots} {2 m \over \hbar^2}
       g_{n+1} | \psi |^{2n} \psi
       \right]
       = 0.
   \label{Schrodinger-GR}
\eea
Using a more proper Klein transformation in Eq.\ (\ref{K-transformation}) for a real field, ignoring the rapidly oscillating terms arising in the \lred{$\phi^{2n + 1}$} term, we simply have an additional equation with the complex conjugation of the above equation for $\psi$. This is the relativistic Schr\"odinger equation written in terms of the wave function $\psi$. Together with the fluid quantities properly constructed, this equation can be combined with Einstein's equation. This will be pursued later \cite{Hwang-Noh-2021}.

The following presentation of the 1PN expression of the Schr\"odinger equation is, in our knowledge, new; a weak gravity expansion attempted in \cite{Chavanis-Matos-2017} is not proper. But, here our purpose is only to show the proper derivation of the non-relativistic Schr\"odinger equation combined with Newtonian gravity in Eqs.\ (\ref{Schrodinger-eq}) and (\ref{Poisson-eq}), or Eqs.\ (\ref{Schrodinger-C}) and (\ref{Poisson-0PN-C}) in the cosmological context. For this pedagogic purpose, we present some details involved in the derivation.

In a spatially flat cosmological background, our metric convention to 1PN order is
\bea
   & & g_{00} = - \left( 1 + 2 {\Phi \over c^2} \right), \quad
       g_{0i} = - a {P_i \over c^3}, \quad
       g_{ij} = a^2 \left( 1 - 2 {\Psi \over c^2} \right) \delta_{ij},
\eea
where the spatial index of $P_i$ is raised and lowered using $\delta_{ij}$ and its inverse; here, index $0 = c t$. In order to properly consider the 1PN expansion, we have to include $c^{-4}$-order in $g_{00}$, see Eq.\ (\ref{Chadrasekhar-notation}); thus, $\Phi$ includes $c^{-2}$ order.
The inverse metric and connection, valid to 1PN order, are
\bea
   & & g^{00} = - \left( 1 - 2 {\Phi \over c^2}
       + 4 {\Phi^2 \over c^4} \right), \quad
       g^{0i} = - {1 \over a} {P^i \over c^3}, \quad
       g^{ij} = {1 \over a^2}
       \left( 1 + 2 {\Psi \over c^2} \right) \delta^{ij},
   \\
   & & \Gamma^0_{00} = {\dot \Phi \over c^3}
       - {1 \over c^5} \left( 2 \Phi \dot \Phi
       + {1 \over a} P^i \Phi_{,i} \right), \quad
       \Gamma^0_{0i} = {\Phi_{,i} \over c^2}
       - {1 \over c^4} \left( 2 \Phi \Phi_{,i}
       + a H P_i \right),
   \nonumber \\
   & &
       \Gamma^0_{ij} = {1 \over c} a^2 H \delta_{ij}
       - {1 \over c^3} a^2 \left[ \dot \Psi
       + 2 H \left( \Phi + \Psi \right) \right] \delta_{ij}
       + {1 \over c^3} a P_{(i,j)}, \quad
       \Gamma^i_{00} = {1 \over a^2} {\Phi^{,i} \over c^2}
       + {1 \over c^4} \left[ 2 \Psi \Phi^{,i}
       - ( a P^i )^{\displaystyle{\cdot}} \right],
   \nonumber \\
   & &
       \Gamma^i_{0j} = {1 \over c} H \delta^i_j
       - {\dot \Psi \over c^3} \delta^i_j
       + {1 \over c^3} {1 \over 2 a} \left( P_j^{\;\;,i} - P^i_{\;\;,j} \right), \quad
       \Gamma^i_{jk} = - {1 \over c^2} \left( \Psi_{,k} \delta^i_j
       + \Psi_{,j} \delta^i_k - \Psi^{,i} \delta_{jk} \right).
\eea
These are the same as Eqs.\ (1)-(5) in \cite{Hwang-Noh-Puetzfeld-2008}. Using these, to 1PN order, Eq.\ (\ref{Schrodinger-GR}) gives
\bea
   & & {2 i m \over \hbar} \left( \dot \psi
       + {3 \over 2} H \psi \right)
       + {\Delta \over a^2} \psi
       - {2 m^2 \over \hbar^2} \Phi \psi
       - \sum_{n = 1, 2 \dots} {2 m \over \hbar^2}
       g_{n+1} | \psi |^{2n} \psi
       + {1 \over c^2} \bigg[
       - \ddot \psi - 3 H \dot \psi
       + 2 \Psi {\Delta \over a^2} \psi
       + {1 \over a^2} \left( \Phi - \Psi \right)^{,i} \psi_{,i}
   \nonumber \\
   & & \qquad
       + {2 i m \over \hbar}
       \left( - 2 \Phi \dot \psi
       + {1 \over a} P^i \psi_{,i} \right)
       + {i m \over \hbar} \left(
       - \dot \Phi - 3 \dot \Psi - 6 H \Phi
       + {1 \over a} P^i_{\;\;,i} \right) \psi
       + {4 m^2 \over \hbar^2} \Phi^2 \psi
       \bigg]
       = 0.
\eea
To the 0PN order ($c \rightarrow \infty$ limit) we recover Eq.\ (\ref{Schrodinger-C}), and Eq.\ (\ref{Schrodinger-eq}) in a static background. The first $\Phi$ term appearing in this equation still contains 1PN order. Using Chandrasekhar's 1PN notation in \cite{Chandrasekhar-1965}
\bea
   & & \Phi = - U + {1 \over c^2}
       \left( U^2 - 2 \Upsilon \right), \quad
       \Psi = - V,
   \label{Chadrasekhar-notation}
\eea
with $V = U$ to 0PN order, we have
\bea
   & &
       {2 i m \over \hbar} \left( \dot \psi
       + {3 \over 2} H \psi \right)
       + {\Delta \over a^2} \psi
       + {2 m^2 \over \hbar^2} U \psi
       - \sum_{n = 1, 2 \dots} {2 m \over \hbar^2}
       g_{n+1} | \psi |^{2n} \psi
       + {1 \over c^2} \bigg[
       - \ddot \psi - 3 H \dot \psi
       - 2 U {\Delta \over a^2} \psi
   \nonumber \\
   & & \qquad
       + {2 i m \over \hbar} \left( 2 U \dot \psi
       + {1 \over a} P^i \psi_{,i} \right)
       + {i m \over \hbar}
       \left( 4 \dot U + 6 H U
       + {1 \over a} P^i_{\;\;,i} \right) \psi
       + {2 m^2 \over \hbar^2}
       \left( U^2 + 2 \Upsilon \right) \psi
       \bigg]
       = 0.
   \label{Schrodinger-1PN}
\eea
To 1PN order, we still have a freedom to impose the temporal gauge (hypersurface or slicing) condition; the spatial gauge condition is already imposed with all the remaining variables spatially gauge invariant, and for various temporal gauge conditions, see Section 6 of \cite{Hwang-Noh-Puetzfeld-2008}. Together with the fluid quantities properly constructed to 1PN order, this can be combined with Einstein's equation expanded to 1PN order \cite{Hwang-Noh-Puetzfeld-2008}.

Now we derive the Poisson equation to 0PN order. To 0PN order, we have
\bea
   & & g_{00} = - \left( 1 + 2 {\Phi \over c^2} \right), \quad
       g_{0i} = 0, \quad
       g_{ij} = a^2 \delta_{ij},
   \label{metric-0PN}
\eea
and no gauge condition is needed. Under the Klein transformation in Eq.\ (\ref{K-transformation}), the energy-momentum tensor in Eqs.\ (\ref{Tab}) and (\ref{V}) gives
\bea
   & & T_{ab} = {\hbar^2 \over m} \bigg\{
       \psi_{,(a} \psi^*_{,b)}
       + {i m c \over \hbar}
       \left( \psi_{,(a} \delta^0_{b)} \psi^*
       - \psi^*_{,(a} \delta^0_{b)} \psi \right)
       + {m^2 c^2 \over \hbar^2} | \psi |^2
       \delta^0_a \delta^0_b
   \nonumber \\
   & & \quad
       - {1 \over 2} \left[ \psi^{;c} \psi^*_{,c}
       + {i m c \over \hbar}
       \left( \psi^{;0} \psi^* - \psi^{*;0} \psi \right)
       + {m^2 c^2 \over \hbar^2} | \psi |^2
       \left( g^{00} + 1 \right)
       + \sum_{n = 1, 2 \dots} {2 m \over \hbar^2}
       {g_{n+1} \over n+1} | \psi |^{2n + 2}
       \right] g_{ab}
       \bigg\},
   \label{Tab-MSF-psi}
\eea
where we used $c t_{,a} = \delta^0_a$, and ignored the rapidly oscillating terms. This is exact. The same result can be derived using the original Klein transformation in Eq.\ (\ref{Klein-transformation}), this time without need for ignoring the oscillating terms, by an appropriate complex conjugation of the energy-momentum tensor, as
\bea
   & & T_{ab}
       = \phi_{,(a} \phi^*_{,b)}
       - \left( {1 \over 2} \phi^{;c} \phi^*_{,c}
       + {1 \over 2} {m^2 c^2 \over \hbar^2} | \phi |^2
       + \sum_{n = 1, 2 \dots} {g_{n+1} \over n+1}
       \left( {m \over \hbar^2} | \phi |^2 \right)^{n+ 1}
       \right) g_{ab},
\eea
which is the same as Eqs.\ (\ref{Tab}) and (\ref{V}) for our real scalar field.

To 0PN (or $c \rightarrow \infty$) limit, we have
\bea
   & & T_{ab} = m c^2 | \psi |^2 \delta^0_a \delta^0_b.
\eea
Thus,
\bea
   & & T_{00} = m c^2 | \psi |^2, \quad
       T_{0i} = 0, \quad
       T_{ij} = 0.
   \label{T_ab-0PN}
\eea
Equation (\ref{Tab-MSF-psi}) shows that {\it only} in the non-relativistic limit, we can identify $m | \psi |^2$ as the mass-density.
Using
\bea
   & & R_{00} = {1 \over c^2}
       \left( - 3 {\ddot a \over a}
       + {\Delta \over a^2} \Phi \right),
\eea
the 00-component of Einstein's equation
\bea
   & & R^a_b = {8 \pi G \over c^4} \left( T^a_b
       - {1 \over 2} T \delta^a_b \right)
       + \Lambda,
\eea
gives
\bea
   & & {\Delta \over a^2} \Phi
       = 4 \pi G m | \psi |^2 + 3 {\ddot a \over a} - \Lambda c^2.
   \label{Poisson-0PN}
\eea
This is Eq.\ (\ref{Poisson-0PN-C}); in a static background, with $a \equiv 1$, we recover Eq.\ (\ref{Poisson-eq}). To the background order, we have
\bea
   & & {\ddot a \over a} = - {4 \pi G \over 3} m | \psi_b |^2
       + {\Lambda c^2 \over 3},
\eea
thus, in the static background, we have $4 \pi G m | \psi_b |^2 = \Lambda c^2$.
Subtracting the background order, we have
\bea
   & & {\Delta \over a^2} \Phi
       = 4 \pi G m \left( | \psi |^2 - | \psi_b |^2 \right).
   \label{Poisson-BG}
\eea
This avoids the so-called Jeans swindle, with Poisson's equation applying {\it only} for inhomogeneous part: i.e., no gravitational potential for the homogeneous background as Jeans has correctly chosen \cite{Jeans-1902}.

\end{widetext}
%
%


\end{document}